\documentclass[manuscript]{aastex}
\begin{document}
\title{Scaling the smoothness of the IGM\footnote{Some of the data presented herein were obtained at the W.M. Keck Observatory, which is operated as a scientific partnership among the California Institute of Technology, the University of California and the National Aeronautics and Space Administration. The Observatory was made possible by the generous financial support of the W.M. Keck Foundation.}}

\author{R. Guimaraes \altaffilmark{1}, M. S. Silva and M. A. Moret}
\affil{Programa de Modelagem Computacional - SENAI - Cimatec, 41650-010 Salvador, Bahia, Brazil}
\email{rguimara@on.br} 
\author{P. Petitjean and E. Rollinde}
\affil{Institut d'Astrophysique de Paris \& Universit\'e Pierre et Marie Curie, 98 bis boulevard d'Arago, 75014 Paris, France}
\email{ppetitje@iap.fr}
\and
\author{S. G. Djorgovski}
\affil{California Institute of Technology, 105-24, Pasadena, CA 91125, USA}
\email{george@astro.caltech.edu}

\begin{abstract}
We use, for the first time, the Detrend Fluctuation Analysis (DFA) to study the correlation properties of the transmitted flux fluctuations, in the Lyman-$\alpha$ (Ly$\alpha$) Forest along the lines of sight (LOS) to QSOs, at different space scales. 
We consider in our analysis the transmitted flux in the intergalactic medium over the redshift range 2 $\leq$ z $\leq$ 4.5 from a sample of 45 high-quality medium resolution (R $\sim$ 4300) quasar spectra obtained with Echelle Spectrograph and Imager (ESI) mounted on the Keck II 10-m telescope, and from a sample of 19 high-quality high resolution (R $\sim$ 50000) quasar spectra obtained with Ultra-Violet and Visible Echelle Spectrograh (UVES) mounted on the ESO KUEYEN 8.2 m telescope. 
The result of the DFA method applied to both datasets, shows that there exists a difference in the correlation properties between the short and long-range regimes: the slopes of the transmitted flux fluctuation function  are different on small and large scales. 
The scaling exponents, $\alpha_{1}$ = 1.635$\pm$ 0.115 and $\alpha_{2}$ = 0.758$\pm$ 0.085 for the ESI/Keck sample and $\alpha_{1}$ = 1.763 $\pm 0.128$ and $\alpha_{2}$ = 0.798 $\pm 0.084$ for the UVES/VLT sample for the short and long range regime respectively.
The transition between the two regims is observed at about $\sim 1.4 h^{-1} Mpc$ (comoving). 
The fact that $\alpha_{1}$ is always larger than $\alpha_{2}$ for each spectrum supports the common view that the Universe is smoother on large scales than on small scales. 
The non  detection of considerable variations in the scaling exponents from LOS to LOS 
confirms that anisotropies cannot be ubiquitous, at least on these scales. 

\end{abstract}

\keywords{quasars: general --- quasars: absorption lines --- statistics: DFA}

\section{Introduction}

In the eighties the numerous hydrogen Lyman-$\alpha$ absorption lines, called the \textit{Ly-$\alpha$ forest}, seen in the spectra of quasars were interpreted as revealing a population of intergalactic clouds (see Sargent et al. 1980). 

The advent of numerical simulations changed the understanding of the low column density Ly-$\alpha$ forest, $N_{HI} \leqslant 10^{14.5} cm^{-2}$, from a population of discrete clouds to a smooth medium with density fluctuations produced by the process of structure formation (Cen et al. 1994; Petitjean et al. 1995; Miralda-Escud\'e et al. 1996; Zhang et al. 1998). This paradigm, confirmed by full hydrodinamical simulations (Cen et al. 1994; Zhang, Anninos \& Norman 1995; Miralda-Escud\'e et al. 1996; Hernquist, Katz \& Weinberg 1996; Wadsley \& Bond 1996; Zhang et al. 1997; Theuns et al. 1998; Machacek et al. 2000; Efstathiou, Schaye \& Theuns 2000), also caused a change in the way we mathematically treat the absorption lines. 
Since then the IGM has become a unique tool to study the evolution of the gas in the Universe (Becker et al. 2011, Bolton et al. 2013) and the large scale structures in the universe (e.g. Croft et al. 2002, Busca et al. 2012) thanks to the dramatic increase in the statistics provided by quasar surveys like the Baryonic Oscillation Spectroscopic Survey from SDSS-III (Eisenstein et al. 2011; P\^aris et al. 2012).
 
Because the gravity of the underlying dark matter dominates, it is often assumed that the IGM constitutes a stochastic field of spatially random fluctuations, since the cosmic mass is randomly distributed.
In this scenarioseveral statistical approaches as the flux decrement (DA), the flux power spectrum, the cumulative distribution function (CDF) and principally the probability distribution of Ly-$\alpha$ pixel  optical depths (PDF), have been used to analyze the observed Ly$\alpha$ forest (Fan et al. 2002; Rollinde et al. 2005; Guimar\~aes et al. 2007; Becker et al. 2007; Viel et al. 2008).

Among the stochastic approaches, the detrended fluctuation analysis (DFA) was proposed (Peng et al. 1994) to analyze long-range power-law correlations in non stationary systems. One advantage of the DFA method is that it allows the long-range power-law correlations in signals with embedded polynomial trends that can mask the true correlations in the fluctuations of a noise signal. The DFA method has been applied to analyze DNA and its evolution (Peng et al. 1992, Peng et al. 1994), file editions in computer diskettes (Zebende et al. 1998), economics (Mantegna \& Stanley 1995; Filho et al. 2008), climate temperature behavior (Talkner \& Weber 2000), phase transition (Zebende et al. 2004), astrophysics sources (Moret et al. 2003, Zebende et al. 2005) and cardiac dynamics (Ivanov et al.1996, Ivanov et al. 1999), among others. 

In this paper we propose the application of the DFA method to investigate the correlation properties presented in the Ly-$\alpha$ forest of quasar spectra. The unidimensional sequence of each spectrum was used to estimate the spatial organization of the IGM (building blocks), which is done by looking at spatial correlations. 
This work is organized as follows: In section II, we describe the data. We present the DFA method in section III and the result of the DFA method applied along the line of sight to quasars in section IV. Finally, we discuss the results and conclude in section V. 

\section{Data}
The observational data used in our analysis were obtained from two different observatories (one in the north and another in the south) and telescopes/instruments. The two observational programs are described below.

\subsection{The ESO Large Program (LP) quasar sample}

The first observational data set used in our analysis was obtained from the Ultra-Violet and Visible Echelle Spectrograph (UVES) mounted on the ESO KUEYEN 8.2 m telescope at the Paranal observatory in the course of the ESO-VLT Large Programme (LP) "Cosmological evolution of the Inter Galactic Medium" (PI Jacqueline Bergeron). 
This programme has been devised to gather a homogeneous sample of echelle spectra of 18 bright QSOs, with uniform spectral coverage, resolution and signal-to-noise ratio suitable for studying the intergalactic medium in the redshift range 1.7 $\leq$ z $\leq$ 4.5. Spectra were obtained in service mode and observations were spread over four periods (two years) during 30 nights under good seeing conditions ( $\leq$ 0.8 arcsec). The spectra have a signal-to-noise ratio of $\sim$ 40 to 80 per pixel and a spectral resolution $\geq$ 45000 in the Ly-$\alpha$ forest region. Details of the data reduction can be found in Chand et al. (2004) and Aracil et al. (2004). In our analysis we have only used the pixels that are located between the Ly-$\alpha$ and the Ly-$\beta$ quasar emission lines.

\subsection{The ESI/KECK quasar sample} 

Medium resolution (R $\sim$ 4300) spectra of all z $\geq$ 3 quasars discovered in the course of the DPOSS (Digital Palomar Observatory Sky Survey (see e.g. Kennefick, Djorgovski \& de Carvalho 1995; Djorgovski et al. 1999, Stern et al. 2000 and the complete listing of QSOs available at http://www.astro.caltech.edu/$\sim$george/z4.qsos) have been obtained with the ESI (Sheinis et al. 2002) mounted on the Keck II 10-m telescope. Signal-to-noise ratio (S/N) is usually larger than 15 per 10 km/$s^{-1}$ pixel. These data have already been used to construct a sample of damped Ly-$\alpha$(DLA) systems at high redshift (Prochaska et al. 2003a; Prochaska, Castro \& Djorgovski 2003b,; Guimar\~aes et al. 2009) and to study the density field around quasars (Guimar\~aes et al. 2007). In total, 95 quasars have been observed, and 45 have been selected (because of their high S/N $\geq$ 25) to be used in our analysis. As for the UVES/VLT sample we have used the pixels that are located between the Ly-$\alpha$ and the Ly-$\beta$ QSO emission lines.

\section{The Detrend Fluctuation analysis - DFA}
\begin{figure}[htb]
       \includegraphics[scale=0.8]{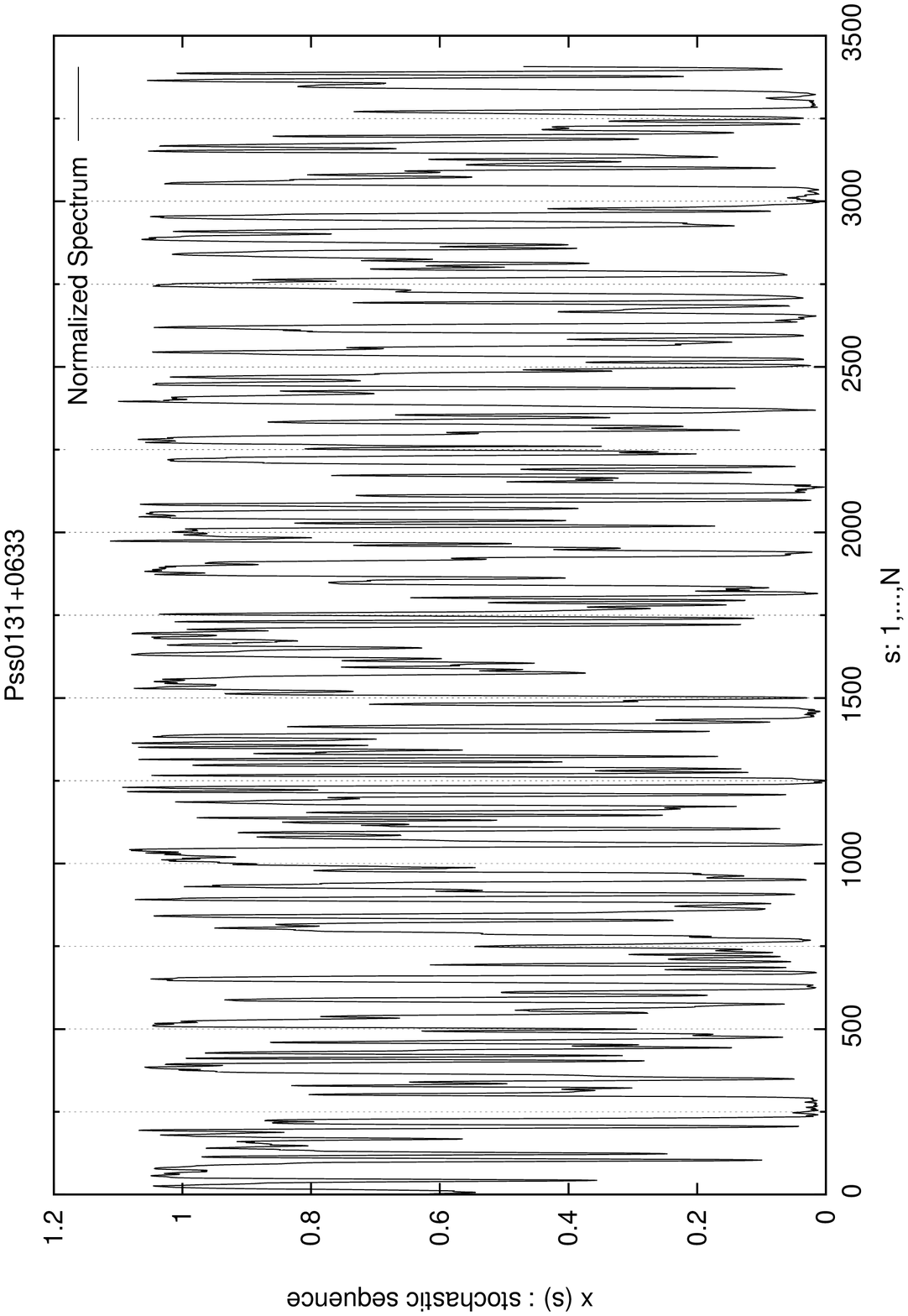}\\
       \caption{ Medium resolution spetrum of the Ly-$\alpha$ forest towards PSS0131+0633 QSO $z_{em}=4.432$, taken with the ESI spectrograph mounted on the Keck II 10-m telescope.}
       \label{spectrum}
\end{figure}

We describe in this section the Detrend Fluctuation Analysis (DFA), wich is designed to treat data sequences with statistical inohomogeneities, such as in quasar spectra arising from non stationarity of the signal. We treat the pixels in the Ly-$\alpha$ forest as a stochastic sequence x(s), s = 1,...,N (see Fig. \ref{spectrum}).
 
\begin{figure}[htb]
       \includegraphics[scale=0.8]{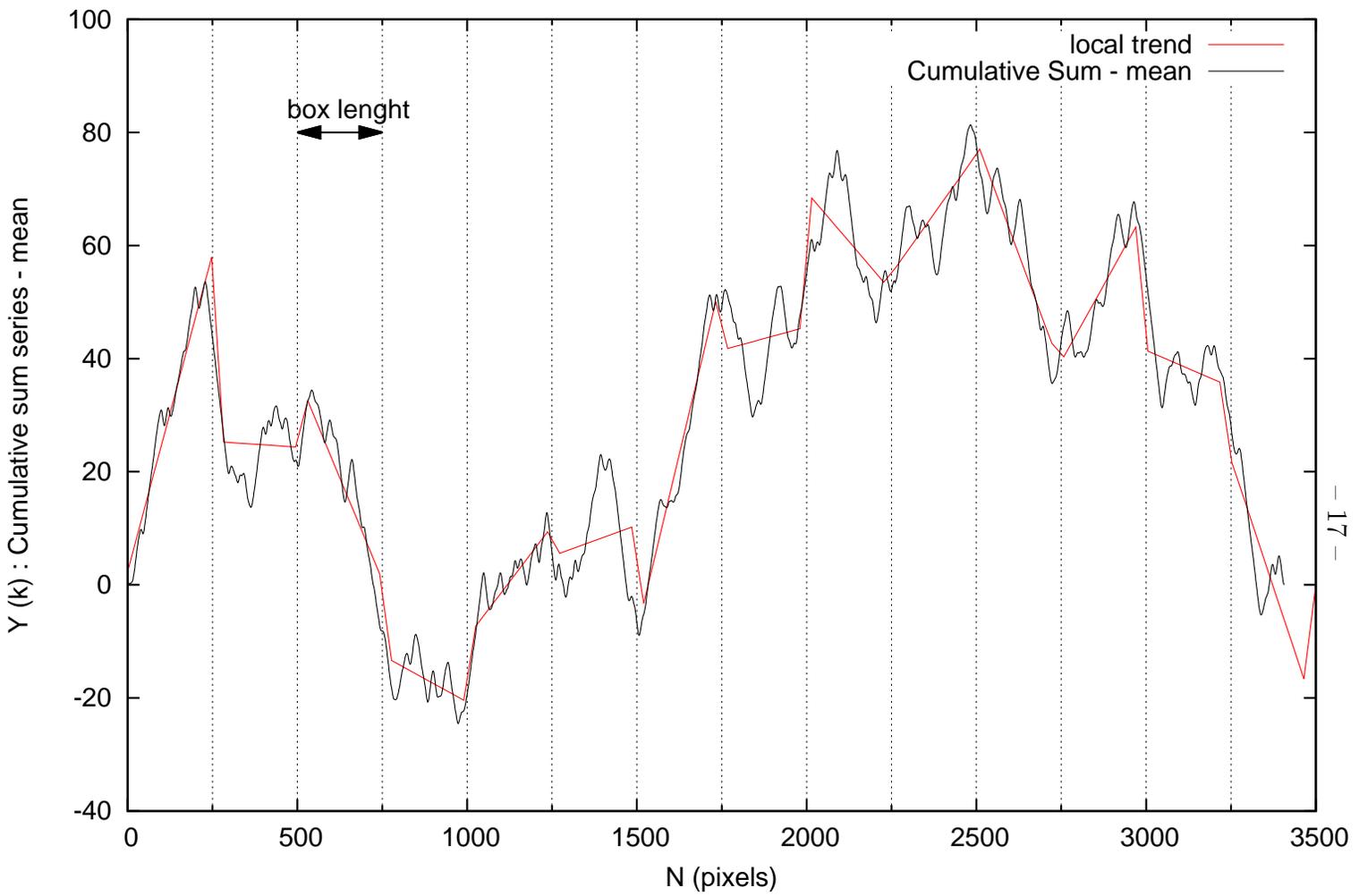}\\
       \caption{ The cumulative sum of pixel series (for PSS0131+0633) after mean subtraction (black line), divided into equally sized box of length 250 (dashed line). A local linear trend was traced (red line) for each box. }
       \label{flutuacao}      
\end{figure}

First, we obtain the cumulative sum of pixel series after subtracted the mean (see Fig \ref{flutuacao}).

\begin{eqnarray}
Y(k) =  \sum_{s=1}^{k} [ x(s) - \bar{x} ] , where\\   
\bar{x} = \frac{1}{N} \sum_{s=1}^{N} x_{s} \nonumber  
\end{eqnarray}
and N is the number of pixels

After that, the profile $Y(k)$ was divided into equally sized box of lenght $n$ (we permit the overlapping of the segments). The range used for the box was $n_{min} \simeq 4$ and $n_{max} \simeq N/3$. For each segment or box we calculated the local trend $y_n(k)$, that represents the non stationarity (linear, quadratic, or polynomial) in that box, by a least square fit of the series. Finally, we subtracted the local trend $y_n(k)$ from the integrated series $Y(k)$ then calculating the root-mean square (RMS) deviation from the trend (fluctuation), i. e., the detrend fluctuation function F(n). 

\begin{equation}
F(n) =  \sqrt{\frac{1}{N} \sum_{k=1}^{N} [ Y(k) - y_n(k) ]^2}   
\end{equation}

For each box, covering the entire analyzed space series, we repeated this procedure, trying to found a relationship between the box length $n$ and the fluctuation $F(n)$. If the data are power-law correlated, $F(n)$ increases with $n$, as a power-law, $F(n) \sim n^{\alpha}$. The scaling exponent $\alpha$, quantifying the degree of the long-range correlations, can be obtained from the slope of a straight line fit to $log[F(n)] \sim \alpha \times log(n)$ on a log-log plot.

The scaling exponent $\alpha$ quantifying the degree of the space scale correlations, can take the following values: uncorrelated signal (random walk) yields $\alpha = 0.5$, antipersistent correlations yields $\alpha \leq 0.5$, and persistent correlations indicating the presence of spatial correlations yields $0.5 \le \alpha \le 1.0$. The values $\alpha = 1.0$ and $\alpha = 1.5$ correspond to $\frac{1}{f}$ noise and Brownian motion, respectively. Long-range negatively correlated fluctuations or changes yields $\alpha \le 1.5$, while the positively correlated fluctuations yields $\alpha \ge 1.5$.

\section{Lyman $\alpha$ Forest space scale}

The spatial density fluctuations in the IGM are nonperiodic and constitute a stochastic field. Among the stochastic approaches, the DFA method described in the previous section is aimed at demonstrating the presence of scale-invariant self-similar features (correlation, memory) in the spatial linear density variations along the LOS to quasars. 

The DFA method was applied to the two available data sets separately.

\subsection{ESI/Keck sample}

\begin{figure}[htb]
           \includegraphics[scale=0.8]{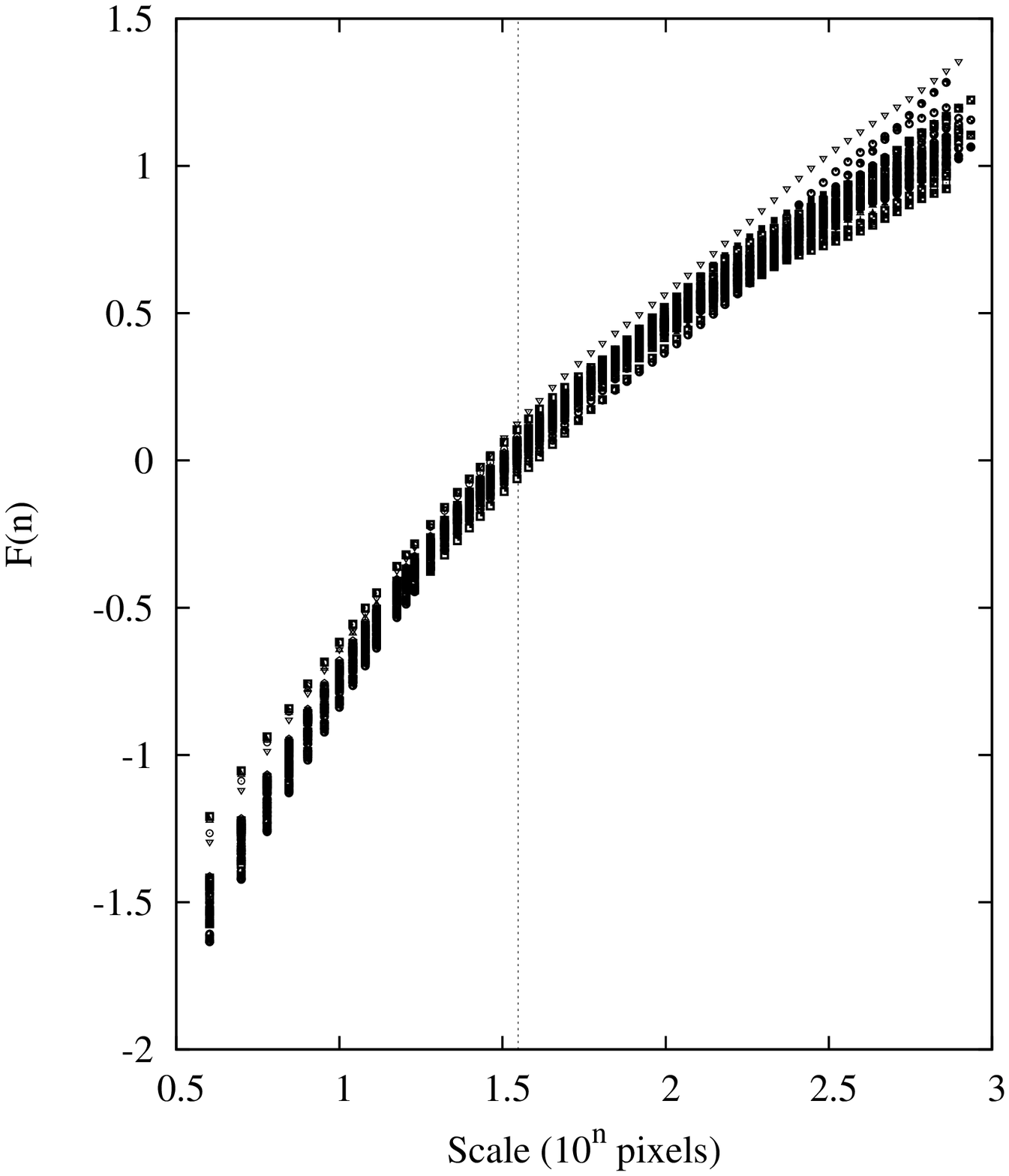}\\
           \caption{The plot shows the transmitted flux fluctuations as a function of pixel scale for the ESI/Keck spectra sample.}
           \label{keck_scale}
\end{figure}

The scaling analysis results for the Ly-$\alpha$ forest of each ESI/Keck spectrum are shown in figure \ref{keck_scale}. Note that the scaling variation is not the same on small and large scales. We find that for all sequences this line was not straight but has a changing slope.
We determine 2 scaling exponents, $\alpha_{1} = 1.64 \pm 0.12$ and $\alpha_{2} = 0.76 \pm 0.08$ for the short-range and long-range correlations.
 
The average of all crossover points for the Keck data occurs at about $N_{turn} = 10^{1.55 \pm 0.11} = 35^{+10}_{-7}$ pixels. The pixels do not have a constant size in Mpc, so $35^{+10}_{-7}$ pixels can not  match a single distance but is in the range $1.1 - 1.6 h^{-1} Mpc$ (comoving) with a mean value of $\sim 1.3 h^{-1} Mpc$ that is used hereafter.

As $\alpha_{1}$ is always larger than $\alpha_{2}$ we do see stronger correlations on short average space scales up to $\sim 1.3 h^{-1} Mpc$.
 
We can infer from figure \ref{keck_scale} that the fluctuation function presents the same pattern among the different lines-of-sight. 
The scaling exponent $\alpha$ does not vary significantly for different lines-of-sight. This can be considered as a confirmation about the isotropy of the IGM on large scales.

Regarding the variation of $\alpha$ with respect to scale, in figure \ref{keck_scale} we can see that for scales n below $\sim 35$ pixels the scaling exponent is larger than 1.5, reflecting highly deterministic correlations that can be related to the "jeans lenght" (see Bi et al. 1992; Fang et al. 1993; Nusser \& Haehnelt 1999; Nusser 2000). The scaling exponent decays with respect to scale n, approaching an asymptotic value of $\sim 0.65$ and is characterized by the positive long-range weak correlation/persistency that can be related to the skeletonized weak clustering present in the IGM on large scales.

\begin{figure}[htb] 
   \includegraphics[scale=0.8]{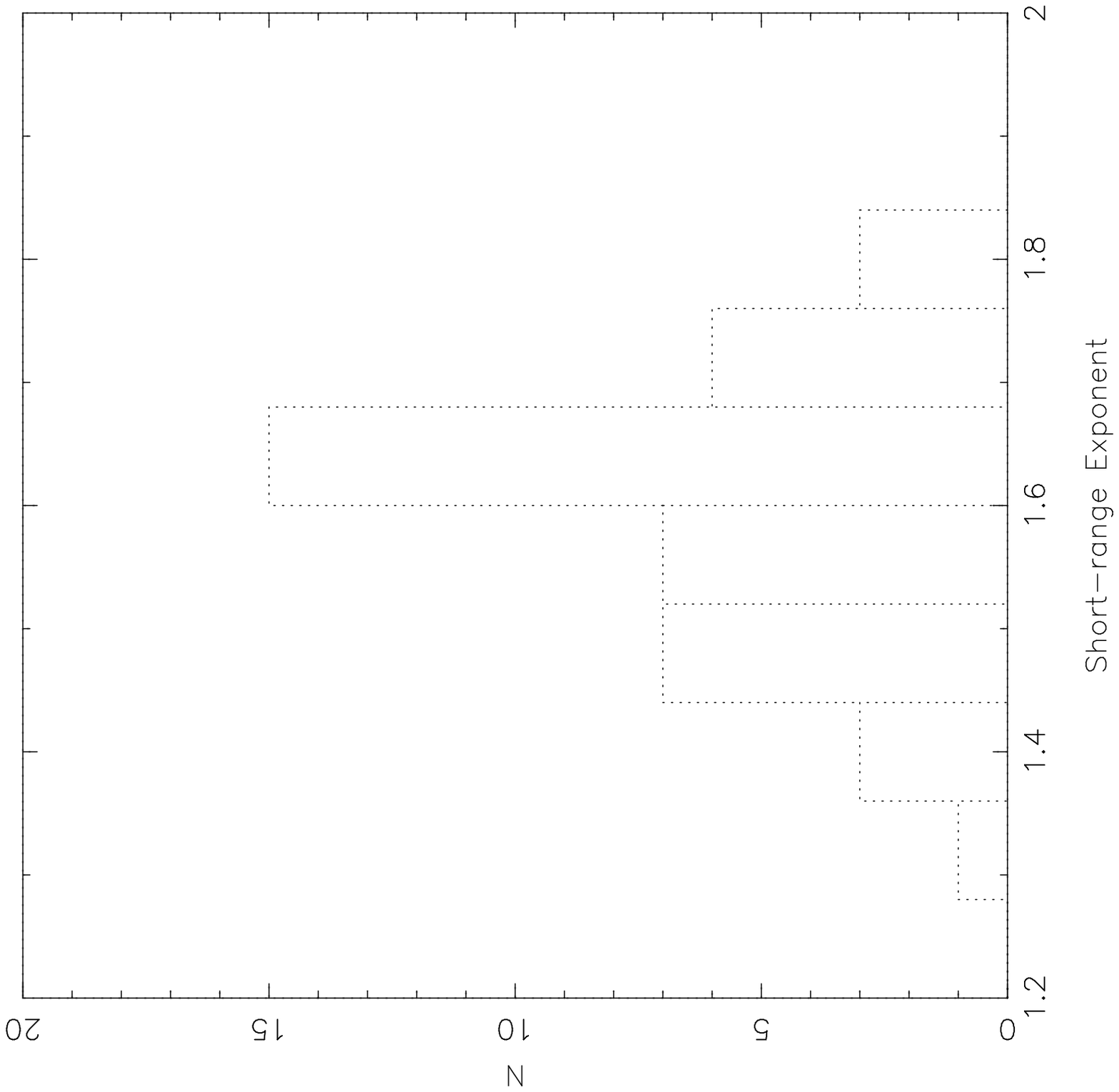}\\
   \caption{Scaling exponent histogram for the short-range scale.}
   \label{keck_short}
\end{figure}

\begin{figure}
   \includegraphics[scale=0.8]{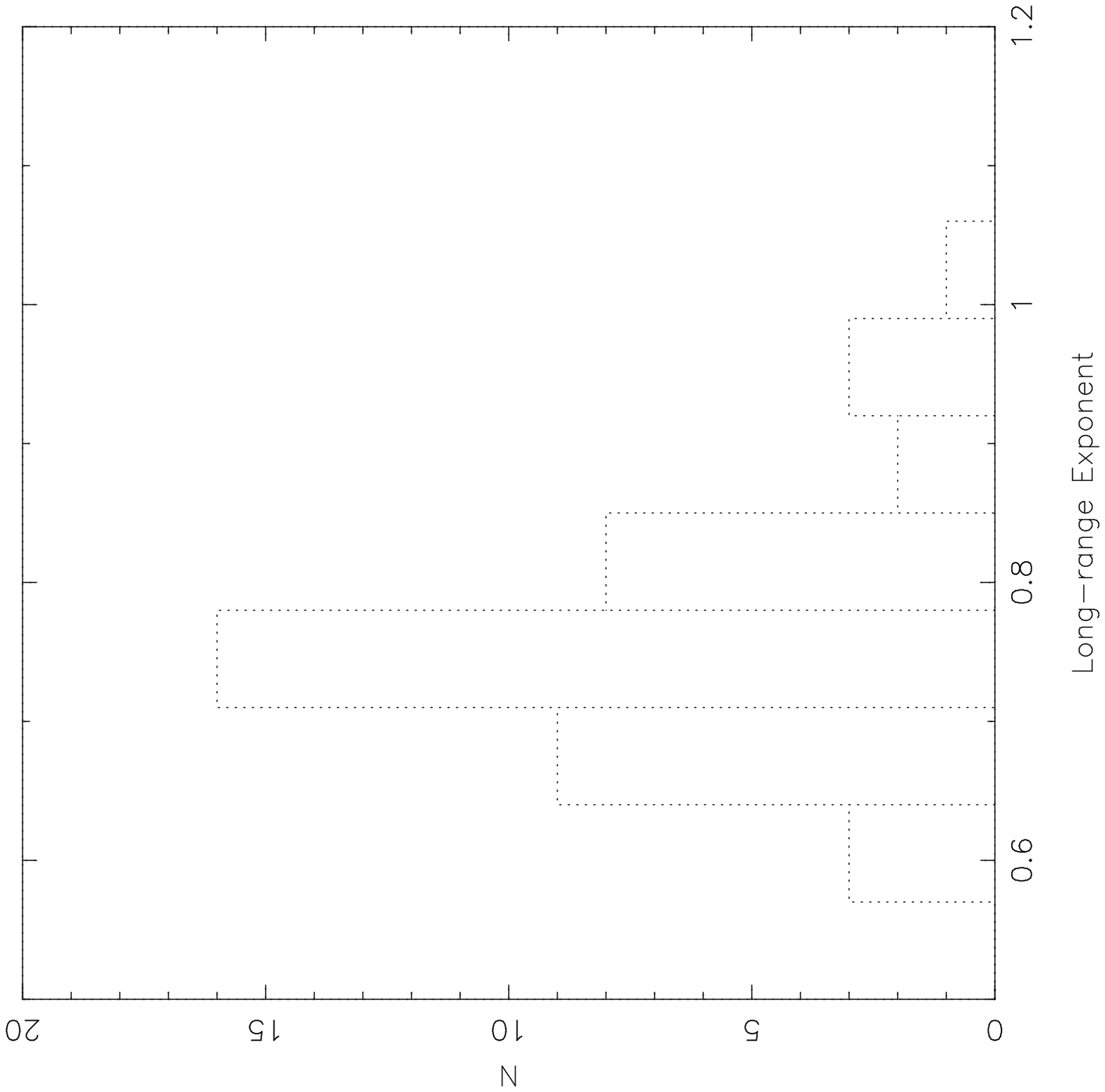}\\
   \caption{Scaling exponent histogram for the long-range scale.}
   \label{keck_long}
\end{figure}

In Fig.~\ref{keck_short} we show the scaling exponent histogram for the short-range regime and in  Fig.~\ref{keck_long} the histogram for the long-range regime.
We can also convert the $\alpha$ scaling exponent, in a fractal dimension that generalises our intuitive concepts of dimension.

\begin{equation}
D = 3 - \alpha
\end{equation} 

We obtained for scales below $\sim 1.3 h^{-1} Mpc$ (comoving) an average fractal dimension value of $D = 1.4$ and for scales above $\sim 1.3 h^{-1} Mpc$ until $\sim 500 h^{-1} Mpc$ (comoving)  an average fractal dimension value of $D = 2.3$. 
Cold Dark Matter models of density fluctuations predict that at scales below $\sim 10 h^ {-1} Mpc$ we detect values of D smaller than 2, with $D \sim 3$ on scales larger than $\sim 100 h^ {-1} Mpc$ (de Gouveia dal Pino et al. 1995; de Vega et al. 1998; Tatekawa \& Maeda 2001 ).

\subsection{The UVES ESO-LP sample}

\begin{figure}[htb] 
           \includegraphics[scale=0.8]{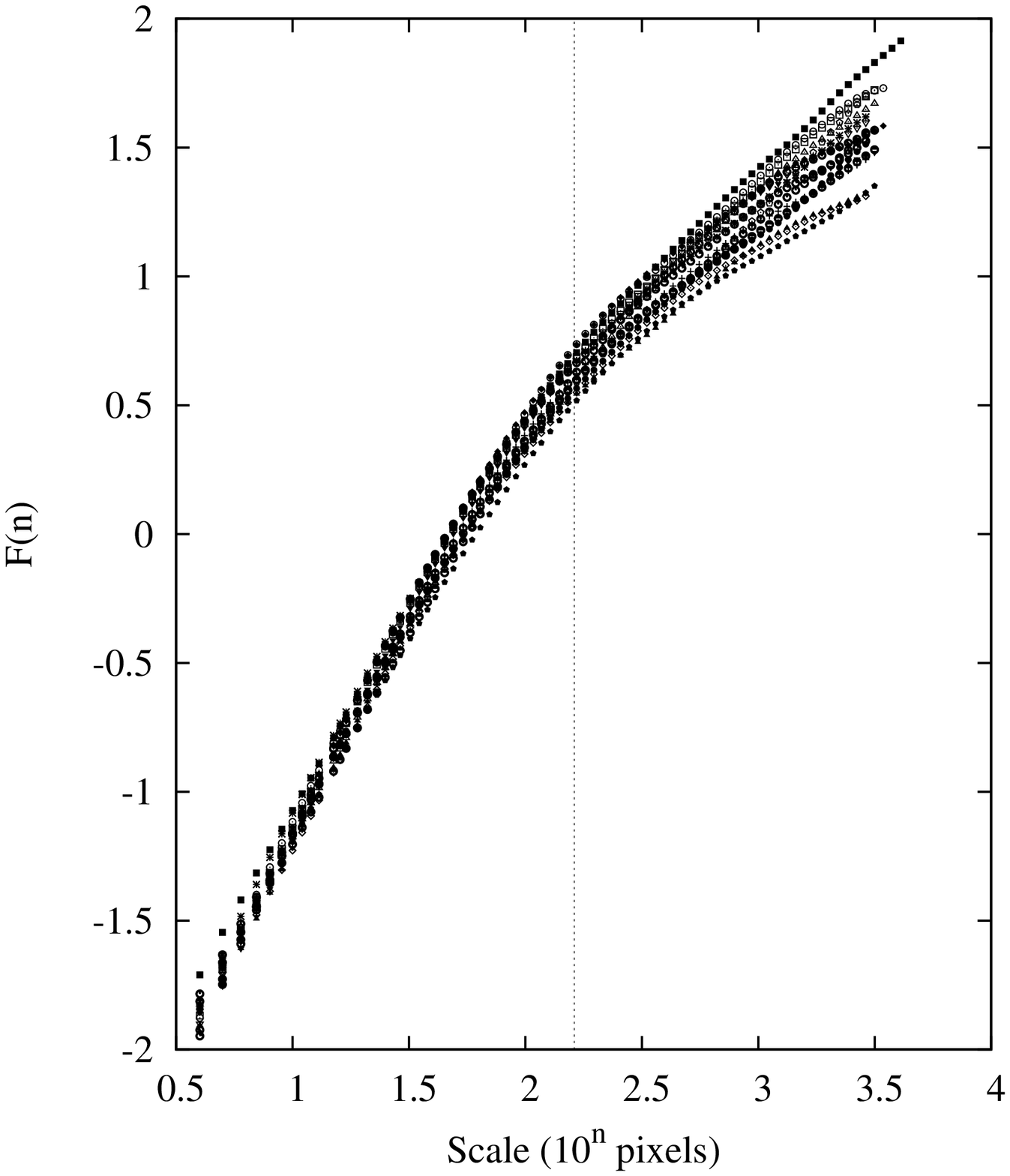}\\
           \caption{The plot shows the transmitted flux fluctuations as a function of pixel scale for the UVES/VLT spectra sample.}
           \label{UVES_scale}
\end{figure}

The scaling analysis results for the Lyman-$\alpha$ forest of each UVES/VLT spectrum are shown in figure \ref{UVES_scale}. Note that, like for the ESI/keck data, the scaling variation on small and large scales is not identical. We determine, as for the ESI/Keck sample, 2 scaling exponents, $\alpha_{1} = 1.76 \pm 0.13$ and $\alpha_{2} = 0.80 \pm 0.08$ for the short-range and long-range correlations, respectively, as the average of all scaling exponents obtained from our sample of 19 spectra. 

The crossover point for the UVES data occurs at about $N_{turn} = 10^{2.207 \pm 0.096} = 161 \pm 4$ pixels. However as for the ESI/Keck sample the pixels do not have a constant size in Mpc, so $161 \pm 4$ pixels can not  match a single distance but at a range of distances $1.0 - 1.8 h^{-1} Mpc$ (comoving).
Note that the spectral resolution is not the same for both instruments ESI/Keck - UVES/VLT which means that pixel sizes are different. The range of turn-over distances are however nearly the same. 

\begin{figure}[htb]
       \includegraphics[scale=0.8]{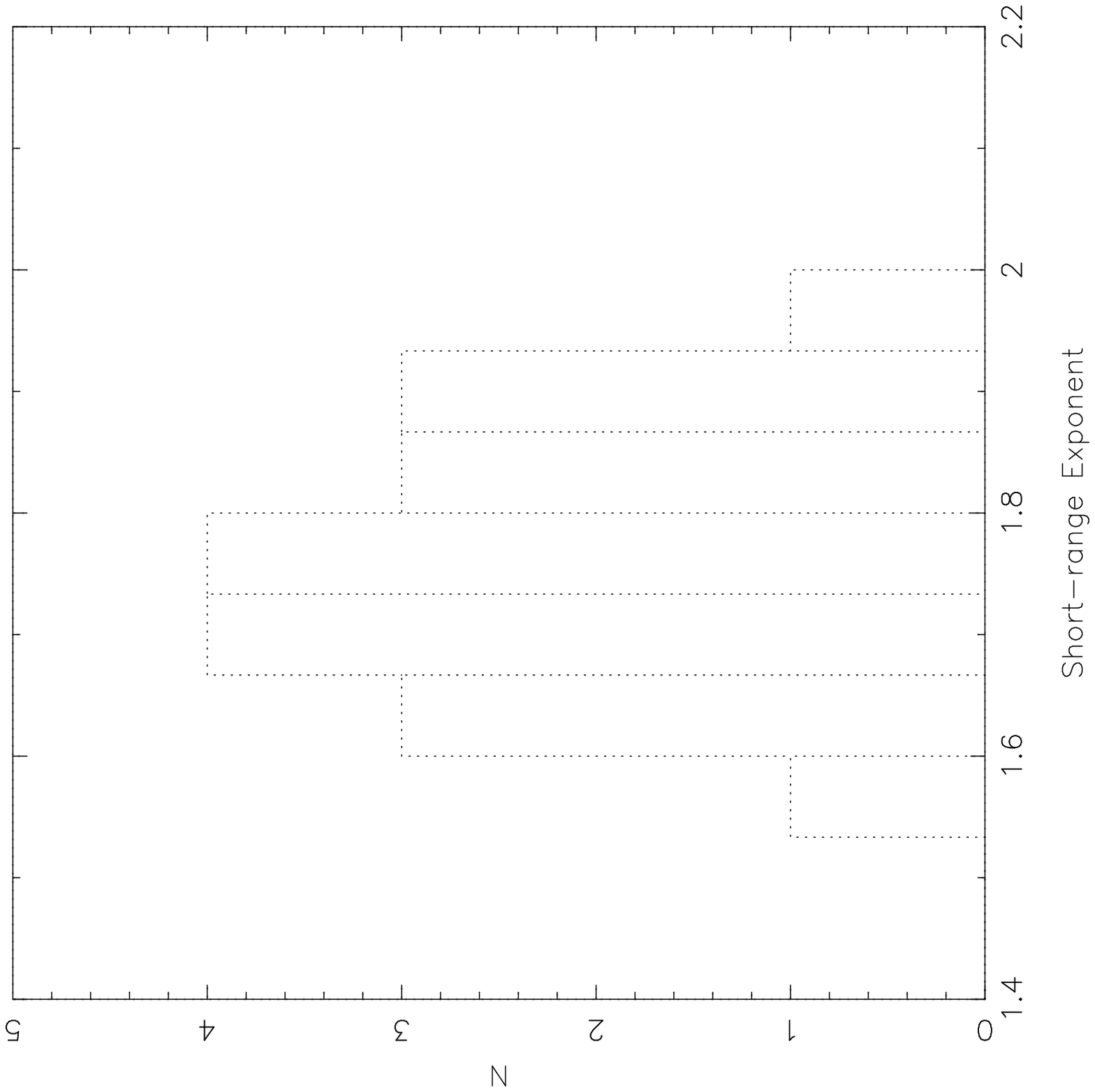}\\
       \caption{Scaling exponent histogram for the short-range scale.}
       \label{uves_short}
\end{figure}

\begin{figure}
	\includegraphics[scale=0.8]{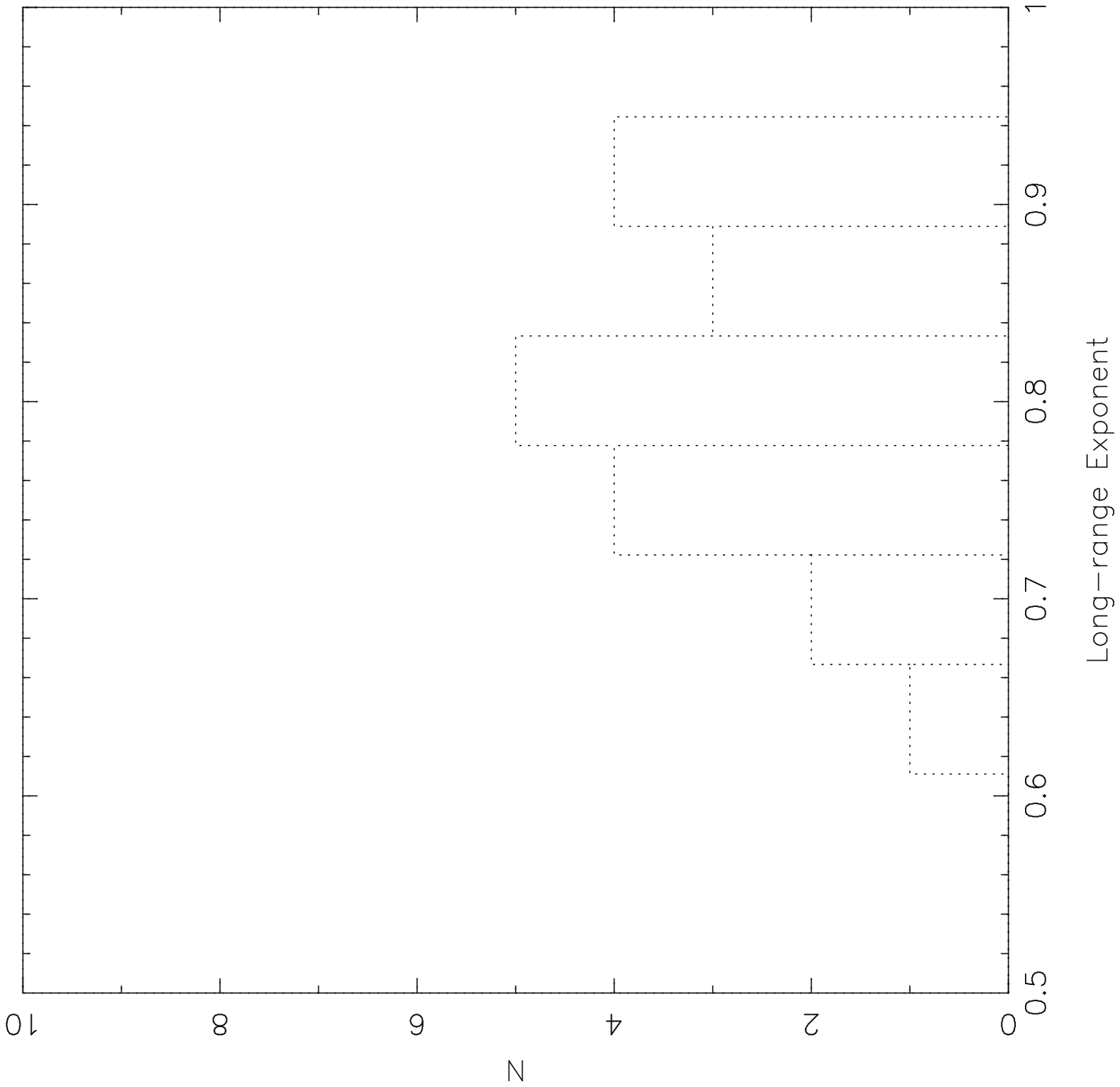}\\
    \caption{Scaling exponent histogram for the long-range scale.}
    \label{uves_long}
\end{figure}

As for the ESI/Keck sample, $\alpha_{1}$ is always larger than $\alpha_{2}$ implying stronger correlations on short average space scales up to $1.4 h^{-1} Mpc$ (comoving).
In Fig.~\ref{uves_short} we show the scaling exponent histogram for the short-range regime and in Fig.~\ref{uves_long} the histogram for the long-range regime.

\section{Conclusion}

The present study investigates fluctuations in the density field of the IGM using an extended random walk analysis, referred to as DFA. 
The advantage of the DFA method is that it can more accurately quantify the correlation property of original signals, even if masked by nonstationarity (in the form of the trends) compared with traditional methods such as flux decrement, power spectrum analysis, cumulative distribution function and probability distribution function. Futhermore the DFA is not sensitive to continuous level uncertainties.

The DFA plots, that is, plots of $log_{10} F(n)$ vs $log_{10} n$ exhibit the so-called "cross-over" phenomenon: 
there exists a difference in the correlation properties between the short and long-range regimes. These $\alpha 1$ and $\alpha 2$ as well as the crossover point, were determined for 2 different sets of data - ESI/Keck UVES/VLT - by the best two-line fit based on least $\chi$ squared test.

The small scale scaling exponent estimated from the ESI/Keck set of data, $\alpha_{1} = 1.635 \pm 0.115$, and from the UVES/VLT set of data, $1.763 \pm 0.128$, are larger than 1.5 indicating positive strong spatial autocorrelation or persistency in the subsequent increases or decreases. 
In the long-range regime the scaling exponents for the set of ESI/Keck data is $\alpha_{2} = 0.758 \pm 0.085$ and that from the UVES/VLT set of data is $\alpha_{2} = 0.798 \pm 0.084$. This implies that on longer scales we found a weaker positive correlation or persistency than on short range scales. 
We can interpret this transition with scale as a transition from clumpiness to homogeneity. The scaling exponent is identical in all directions which demonstrate the isotropy (self-similarity) of the IGM.
  
The transition between short and long range scales is found to be at $\sim 1.4 h^{-1}$ Mpc (comoving) and could be related somehow to the comoving Jeans length, that is $\sim 1.0 h^{-1}$ Mpc (comoving) at around z = 3. 
 
The aggreement between the results obtained on two independent sets of data, ESI/Keck and UVES/VLT, increases the confidence in the reliability of our conclusions.

\begin{acknowledgments}
This work received financial support from FAPESB (organization of the Bahia, Brazil, government devoted to funding of science and technology).
SGD was supported in part by the NSF grants AST-0407448, AST-0909182, and AST-1313422.
The authors wish to recognize and acknowledge the very significant cultural role and reverence that the summit of Mauna Kea has always had within the indigenous Hawaiian community. We are most fortunate to have the opportunity to conduct observations from this mountain. We acknowledge the Keck support staff for their efforts in performing these observations.
\end{acknowledgments}

\newpage

\end{document}